\documentclass[11pt]{article}
\usepackage[utf8x]{inputenc}
\usepackage{amssymb,amsmath,mathrsfs,enumerate,mathtools}
\usepackage[compat=1.1.0]{tikz-feynman}
\usepackage{graphicx,rotate,multicol,braket, multirow}
\usepackage{cite}
\usepackage{braket}
\usepackage{float}
\usepackage{slashed}
\usepackage[normalem]{ulem}
\usepackage{wrapfig}
\usepackage[textfont=it,labelfont=bf]{caption}
\usepackage{bm}
\usepackage[colorlinks=true,
linkcolor=red,
urlcolor=blue,
citecolor=blue]{hyperref}
\usepackage{lineno}
\usepackage{color}
\usepackage{bbold}
\long\def\rpl#1!!#2!!{\textcolor{red}{#1} \textcolor{blue}{#2}}

\def\bar{\overline}

\evensidemargin=\oddsidemargin
\def\Eqn#1{Eq.\ (\ref{#1})}
\def\Eqs#1#2{Eqs.\ (\ref{#1}) and (\ref{#2})}

\allowdisplaybreaks

\title{\Large\bf 
	Sign of the $hZZ$ coupling and implication for new physics
}

\author{
	\sf 
	Dipankar Das$^{a,}$\footnote{d.das@iiti.ac.in},
	Anirban Kundu$^{b,}$\footnote{anirban.kundu.cu@gmail.com},
	Miguel Levy$^{c,}$\footnote{miguelplevy@tecnico.ulisboa.pt},
	Anugrah M. Prasad$^{a,}$\footnote{anugrahmprasad@gmail.com},
	Ipsita  Saha$^{d,}$\footnote{ipsita@iitm.ac.in},
	Agnivo Sarkar$^{e,}$\footnote{agnivosarkar@hri.res.in}
	\\[3mm]
	\small\em
	$^a$Indian Institute of Technology (Indore), Khandwa Road, Simrol,
	Indore 453 552, India \\
	\small\em
	$^b$ Department of Physics, University of Calcutta, 92 Acharya Prafulla Chandra Road, Kolkata 700009, India\\
	\small\em
	$^c$ Centro de F\'isica Te\'orica de Part\'iculas-CFTP and Departamento de
	F\'isica,  Instituto Superior T\'ecnico,\\
	\small\em
	Universidade de Lisboa, Av
	Rovisco Pais, 1, P-1049-001 Lisboa, Portugal \\ 
	\small\em
	$^d$Department of Physics, Indian Institute of Technology Madras, Chennai 600036, India\\
	\small\em
	$^e$Regional Centre for Accelerator-based Particle Physics, Harish-Chandra Research Institute,\\
	\small\em
	 HBNI, Chhatnag Road, Jhunsi, Prayagraj (Allahabad) 211019, India
}

\date{}
\begin{document}
\begin{flushright}
	\small{HRI-RECAPP-2024-01}
\end{flushright} 

{\let\newpage\relax\maketitle}	

%
\renewcommand*{\thefootnote}{\arabic{footnote}}
\setcounter{footnote}{0} 

\begin{abstract}
	The magnitudes of the couplings of the scalar resonance at 125 GeV with the
	SM particles are found to be consistent with those of the SM
	Higgs boson. However, the signs are not experimentally determined in most of the
	cases, a prime example being that with the $Z$-boson pair. In other words, $\kappa_Z^h$,
	the ratio of the couplings of the actual 125 GeV resonance with $ZZ$ and that of
	the SM Higgs boson with the same, is consistent with both $+1$ and $-1$, the latter
	being the `wrong-sign'.
	We argue that the wrong-sign $hZZ$ coupling 
	will necessitate the intervention of new physics below $\mathcal{O}\left(620\right)$~GeV 
	to safeguard the underlying theory from unitarity violation. The strength of the new nonstandard couplings can
be derived from the unitarity sum rules, which are comparable to the SM-Higgs
couplings in magnitude. 
Thus the strong limits from the direct searches at the LHC can help us rule out
the existence of such nonstandard particles with unusually large couplings thereby
disfavoring the possibility of a wrong-sign $hZZ$ coupling.
\end{abstract}

After the discovery of a scalar resonance at the LHC in 2012, its resemblance
to the Standard Model~(SM) Higgs scalar has been under constant experimental
scrutiny. In fact, such investigations constitute a major part of the current
as well as future Higgs precision studies. The information about the Higgs couplings
is presented in the form of Higgs signal-strengths which, in most cases, depend
on the absolute values of the tree-level Higgs couplings. As a result, the information
on the sign of the tree-level Higgs couplings gets washed away except for the
relative sign between the $\bar{t}th$ and $WWh$ ($h$ is the Higgs-like scalar observed at the
LHC) couplings, which can be inferred from the measurement of the $h\to\gamma\gamma$
signal strength. The sign of all the other tree-level Higgs couplings remain
elusive to this day and calls for innovative strategies to probe them.

In this article we embark on an effort to systematically resolve the relative
sign between the $hWW$ and $hZZ$ couplings. To be more specific,
we define the Higgs coupling modifiers as
\begin{eqnarray}
\label{e:kap}
	\kappa_X^h = \frac{g_{XXh}}{g^{\rm SM}_{XXh}} \,,
\end{eqnarray}
where `$X$' is a generic symbol for the SM fermions and massive vector bosons
and $g_{XXh}$ represents the strength of the trilinear coupling at the tree-level.
Here, $h$ in the denominator denotes the SM Higgs boson, and $h$ in the numerator 
denotes $h_{125}$, the observed resonance at 125 GeV. If $\kappa_X^h=1$ for all $X$, we can identify 
$h_{125}$ with $h$.
Although the ratio of $\kappa_W^h$ and $\kappa_Z^h$ is usually linked to the
electroweak $\rho$-parameter due to the custodial symmetry, there exist
models where this ratio can become disentangled from the 
$\rho$-parameter\cite{Chiang:2018irv}.
As a result, it is possible to accommodate $\kappa_W^h= -\kappa_Z^h =1$ while
	keeping $\rho=1$ at the tree-level.
Because of the existence of such models, it becomes imperative to build a strategy
to experimentally probe the scenario with $\kappa_W^h= -\kappa_Z^h =1$. In this context
it was suggested\cite{Chiang:2018fqf,Hamdellou:2021hhr} that one should look into process like 
$e^+e^- \to W^+W^-h$\cite{Bi:2018ffv} which
will involve an interference between the $hWW$ and $hZZ$ vertices 
allowing us, in principle,
to experimentally determine the relative sign between $\kappa_W^h$
and $\kappa_Z^h$. A similar strategy was adopted by the ATLAS and CMS 
Collaborations\cite{CMS:2023sdc,ATLAS:2024vxc},
where the $Wh$ production in the vector boson fusion~(VBF) channel\cite{Stolarski:2020qim} was analyzed and
subsequently it was concluded that the `wrong-sign' limit with $\kappa_W^h= -\kappa_Z^h =1$
is excluded with significance greater than $8\sigma$. Such conclusions, however, come
with an important caveat which is the underlying assumption that the processes under
investigation are mediated by the SM-particles only. As we shall argue shortly, the
wrong-sign arrangement will ordain the presence of new interactions beyond the
SM~(BSM). 
Such an apprehension should warrant parallel investigations which have the
potential to reveal additional insights into the relative sign between $\kappa_W^h$
and $\kappa_Z^h$.

This is where our current work becomes relevant. Our approach relies on the identification
of the fact that any deviation from the SM couplings will violate 
unitarity\cite{Joglekar:1973hh,Horejsi:1993hz,Bhattacharyya:2012tj}
and therefore will require the intervention of new physics~(NP). The energy scale beyond which unitarity is violated~($\Lambda_{\rm UV}^{\rm max}$)
 may be interpreted as the upper limit on the masses of the
new nonstandard particles required for restoration of unitarity. Moreover, the strengths
of these new interactions cannot be arbitrary as they need to satisfy the unitarity
sum rules\cite{Gunion:1990kf}. Such estimates of the coupling strengths of the new particles
can then be used to place lower bounds on the masses of BSM particles~($\Lambda_{\rm NP}^{\rm min}$) using the data from
direct searches at the LHC. If $\Lambda_{\rm NP}^{\rm min} > \Lambda_{\rm UV}^{\rm max}$
then we may conclude that the NP scenario accommodating the wrong-sign possibility
is ruled out.
Excluding the wrong-sign $\kappa_Z^h$ in this method should be comparatively
more robust because the possible concerns regarding UV-completion have been
taken care of by the unitarity sum rules.

\begin{figure}
\centering
\includegraphics[scale=1.0]{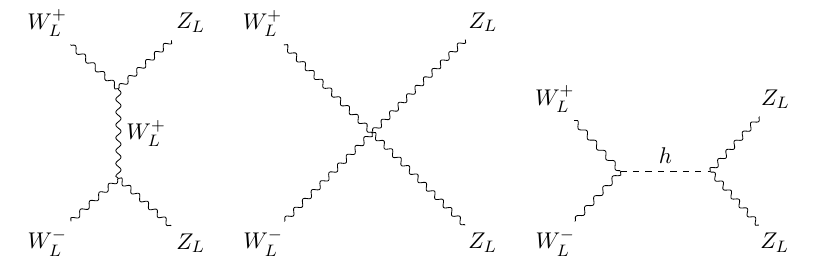}
\caption{The Feynman diagrams for the $W_{L}W_{L} \to Z_{L} Z_{L}$ scattering process which are mediated by the 125 GeV neutral Higgs 
	($h$) and weak vector bosons.}
\label{Fig:WWZZ_SM}
\end{figure}

To illustrate our bottom-up approach explicitly, we first note that in the
limit $\kappa_W^h = - \kappa_Z^h= 1$ the $W_L^+W_L^- \to Z_L Z_L$ (the
subscript `$L$' stands for longitudinal polarization) scattering amplitude
will violate unitarity. This is because the quadratic remnant energy growths from
the gauge diagrams will no longer be canceled by the SM-Higgs mediated
diagrams (see Fig.~\ref{Fig:WWZZ_SM} for the Feynman diagrams) and the 
resulting amplitude will be\cite{Lee:1977eg}:
\begin{eqnarray}
\label{e:WWZZ}
	{\cal M}_{W_LW_L \to Z_LZ_L}\equiv 16\pi a_0 = \frac{g^2E^2}{M_W^2}\left(
	1-\kappa_W^h\kappa_Z^h \right) +{\cal O}\left(1\right) \,,
	\qquad {\rm for} ~~ E\gg M_W \,,
\end{eqnarray}
where $E$ is the CM energy and $a_0$ is the zeroth partial wave amplitude.
Requiring $a_0\le 1$ we may conclude that
unitarity will be violated for energies beyond\footnote{
	If we impose $\left|{\rm Re}(a_0)\right|\le 1/2$, the constraint from unitarity will be even
	more stringent.
}
\begin{eqnarray}
\label{e:UVS}
	\Lambda_{\rm UV}^{\rm max} = \sqrt{\frac{4\pi v^2}{\left| 1-\kappa_W^h \kappa_Z^h\right|}} =
	\sqrt{2\pi}\, v \approx 620~{\rm GeV} \,,
\end{eqnarray}
where in the final steps we set $\kappa_W^h = - \kappa_Z^h= 1$ and 
$v\approx 246$~GeV as the electroweak vacuum expectation value~(VEV).
Therefore, the effects of new BSM interactions must set in before
$\Lambda_{\rm UV}^{\rm max}$. Let us parametrize the set of new couplings
that will, presumably, come to reinstate unitarity, as follows\footnote{
We are considering only scalars as a minimal choice. Inclusion of nonstandard
vector bosons will introduce ${\cal O}\left(E^4 \right)$ energy growths 
in \Eqn{e:WWZZ} leading to more severe consequences for unitarity violation. We have also not considered the charged scalars couplings to SM fermions
	primarily because they will be relevant for processes like $\bar{f}_1f_2
	\to WZ$ which do not involve $\kappa_X^h$. Moreover, as we will argue shortly,
	the most conservative direct search bounds on the nonstandard masses are
obtained in the limit $\delta_k\approx 0$.} 
\begin{eqnarray}
\label{e:LNP}
{\mathscr L}_{\rm NP} &=& \sum_{k=1}^{N_n} H_k \left(\kappa_W^{H_k} gM_W
W^{\mu+}W_\mu^- +\frac{\kappa_Z^{H_k}}{2} \frac{gM_Z}{\cos\theta_w} Z^\mu Z_\mu
-\kappa_f^{H_k} \frac{m_f}{v}\, \bar{f}f \right) \nonumber \\
&& + \left(\sum_{k=1}^{N_c} \delta_k \frac{gM_W}{\cos\theta_w} W_\mu^+Z^\mu H_k^-
 +\sum_{k=1}^{N_d} \xi_k \frac{gM_W}{2} W_\mu^+W^{\mu +} H_k^{--} + {\rm h.c. }\right) \,,
\end{eqnarray}
where $g$ is the $SU(2)_L$ gauge coupling, $\theta_w$ is the weak mixing
angle,  $M_W$ and $M_Z$ are the $W$
and $Z$-boson masses respectively, and $m_f$ is the mass of the fermion,
$f$. The numbers $N_n$, $N_c$ and $N_d$
represent the number of nonstandard neutral, singly-charged and
doubly-charged scalars respectively. From $WW\to WW$ and $WW\to ZZ$
scattering amplitudes we get the following sum rules:
\begin{subequations}
\label{e:VVVV}
\begin{eqnarray}
	\label{e:sumWWWW}
	1 -\left(\kappa_W^h \right)^2 -\sum_{k=1}^{N_n} \left(\kappa_W^{H_k} \right)^2
	+\sum_{k=1}^{N_d} \xi_k^2 &=& 0 \,, \\
	\label{e:sumWWZZ}
	1 -\kappa_W^h \kappa_Z^h  -\sum_{k=1}^{N_n} \kappa_W^{H_k} \kappa_Z^{H_k}
	+\sum_{k=1}^{N_c} \delta_k^2 &=& 0 \,.
\end{eqnarray}
\end{subequations} 
Similarly, from $\bar{f}f\to WW$ and $\bar{f}f\to ZZ$ amplitudes we find
\begin{subequations}
\label{e:ffVV}
	\begin{eqnarray}
	\label{e:sumffWW}
	1 -\kappa_f^h \kappa_W^h  -\sum_{k=1}^{N_n} \kappa_f^{H_k} \kappa_W^{H_k}
	&=& 0 \,, \\
	\label{e:sumffZZ}
	1 -\kappa_f^h \kappa_Z^h  -\sum_{k=1}^{N_n} \kappa_f^{H_k} \kappa_Z^{H_k}
	&=& 0 \,.
	\end{eqnarray}
\end{subequations} 

To get an intuitive sense of the upcoming phenomenological analysis and our
choices for the nonstandard parameters, we first draw the readers' attention
to the sum rule of \Eqn{e:sumWWZZ} in particular, which, in the limit
$\kappa_W^h = - \kappa_Z^h= 1$, reduces to
\begin{eqnarray}
\label{e:delta}
\sum_{k=1}^{N_n} \kappa_W^{H_k} \kappa_Z^{H_k} = 2 + \sum_{k=1}^{N_c} \delta_k^2 \,.
\end{eqnarray}
We should now recall that the new scalars should be allowed by the direct
search constraints to appear before $\Lambda_{\rm UV}^{\rm max}$. The search
channel relevant in this context will be the VBF production
of neutral scalars with subsequent decays into massive vector boson 
pairs\cite{ATLAS:2020tlo,ATLAS:2020fry,ATLAS:2022qlc,ATLAS:2023alo}. 
Our objective will be to dilute the constraint from this
search channel and, for that, we must try to reduce the magnitude of the product
$\kappa_W^{H_k} \kappa_Z^{H_k}$ without compromising the unitarity sum rules.
Thus, in \Eqn{e:delta}, having $\delta_k\approx 0$ would lead to the
least stringent bounds from direct searches and
in what follows we will work under this assumption.

We can follow a similar approach for the sum rule of \Eqn{e:sumWWWW} which, in
the wrong-sign limit, takes the following simpler form
\begin{eqnarray}
\label{e:xi}
\sum_{k=1}^{N_n} \left(\kappa_W^{H_k} \right)^2 = \sum_{k=1}^{N_d} \xi_k^2 \,.
\end{eqnarray}
Thus, the effect of $\kappa_W^{H_k}$ will also propagate into $\xi_k$ which will make
the doubly-charged scalars receive strong constraints from $pp\to H^{++}\to
 W^+W^+$\cite{CMS:2021wlt}. Therefore, one would try to bring down the magnitude of
$\kappa_W^{H_k}$ but we need to be careful so that the value of $\kappa_Z^{H_k}$
does not become too large because of \Eqn{e:delta}. In what follows, we intend to
verify that from all these considerations, the lower limits from direct searches
always lie {\em above} the upper limit set by unitarity. In this way we shall
conclude that non-observation of new BSM resonances in direct searches at the LHC
together with the constraints from unitarity can potentially rule out the 
wrong-sign scenario with  $\kappa_W^h = - \kappa_Z^h= 1$.

\begin{figure}[htbp!]
	\centering
	\includegraphics[width=0.5\textwidth]{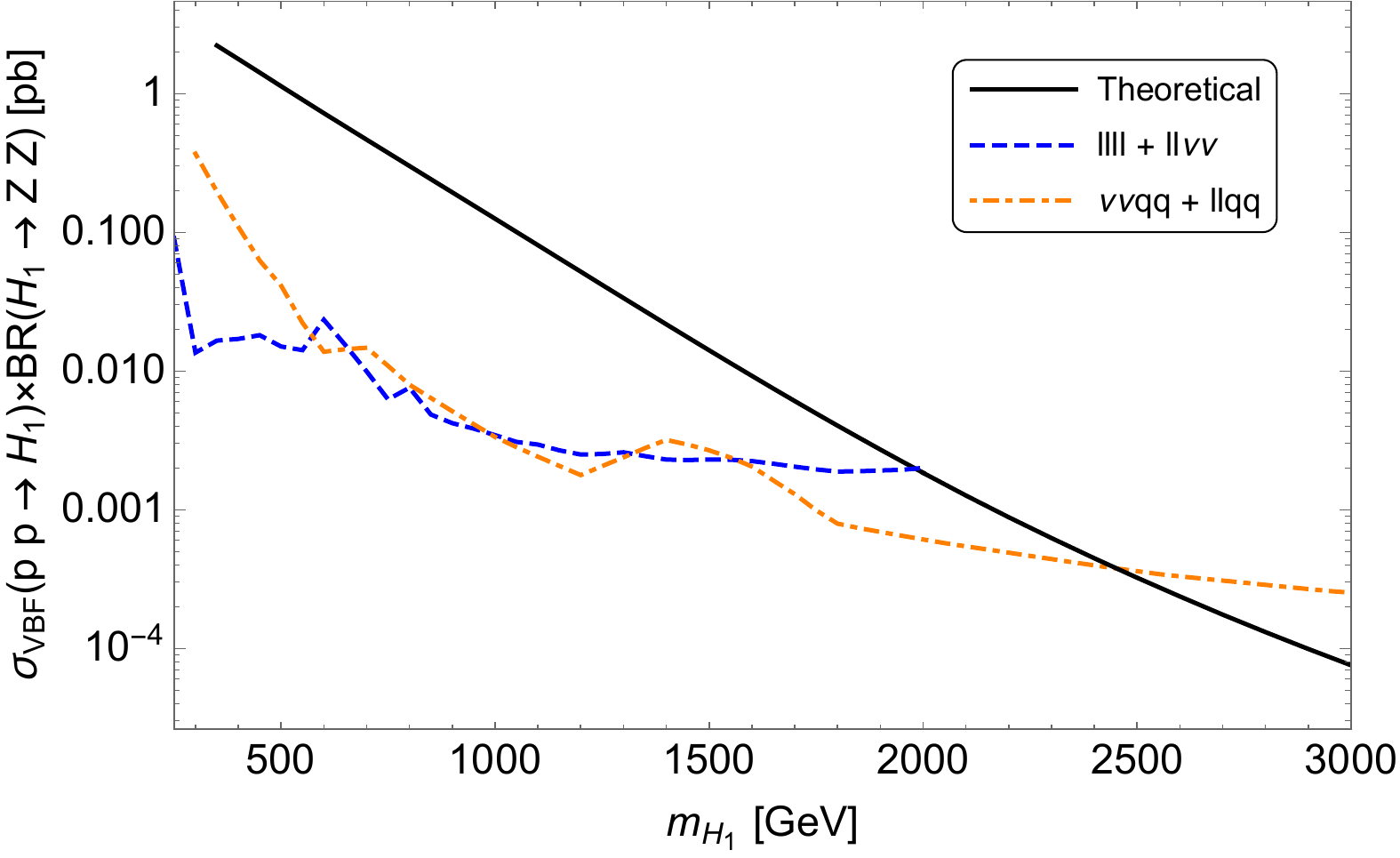}
	\caption{\it The direct search bound on $m_{H_1}$ arising from $pp \xrightarrow{VBF}H_1 \to ZZ$
		for coupling strength given by \Eqn{e:kapZH1}. The blue and orange lines represent the experimental upper limits\cite{ATLAS:2023alo}. 
		To draw the solid black curve (theoretical cross section using the coupling strength
		given by \Eqn{e:kapZH1}) we used the FeynRules \cite{Alloul:2013bka} and the MadGraph \cite{Alwall:2014hca} packages.
		}
	\label{f:1tier}
\end{figure}

%
To begin with, let us investigate whether it is possible to accommodate the wrong-sign
possibility with just one tier of BSM scalars ($N_{n,c,d} =1$). Working under the assumption $\delta_1
\approx 0$, the sum rules of \Eqs{e:VVVV}{e:ffVV}, in the limit $\kappa_W^h =
\kappa_f^h = - \kappa_Z^h= 1$, become
\begin{subequations}
\label{e:sum1}
\begin{eqnarray}
	\kappa_W^{H_1} \kappa_Z^{H_1} &=& 2 \,, \\
	\left(\kappa_W^{H_1}\right)^2 &=& \xi_1^2 \,, \\
	\kappa_W^{H_1}\kappa_f^{H_1} &=& 0 \,, \\
	\kappa_Z^{H_1}\kappa_f^{H_1} &=& 2 \,.
\end{eqnarray}
\end{subequations}
Quite clearly, the above relations cannot be satisfied simultaneously for finite values
of $\kappa_Z^{H_1}$. As the next step, one may try to leverage the experimental
uncertainties in $\kappa_W^h$ and $\kappa_f^h$ (denoted by $\epsilon_W$ and
$\epsilon_f$ respectively)
to accommodate the $\kappa_Z^h= -1$ possibility
and obtain a finite value for $\kappa_Z^{H_1}$. Using $\kappa_W^h = 1\pm
\epsilon_W$, $\kappa_f^h = 1\pm \epsilon_f$ and $\kappa_Z^h=-1$ in the sum
rules of \Eqs{e:VVVV}{e:ffVV} we will obtain the minimum possible value
of $\kappa_Z^{H_1}$ as follows
\begin{eqnarray}
	\label{e:kapZH1}
	\kappa_Z^{H_1} \approx \frac{2}{\sqrt{(\epsilon_W+\epsilon_f)}} \,.
\end{eqnarray}
Current $2\sigma$ limits on $\epsilon_W$ and $\epsilon_f$ stand at $14\%$ and
$19\%$ respectively\cite{CMS:2022dwd}. Using these numbers, the most conservative lower
limit on $m_{H_1}$, from $pp \xrightarrow{VBF}H_1 \to ZZ$ searches as shown in
Fig.~\ref{f:1tier}, is found to be around $2.5$~TeV.\footnote{Here we are implicitly 
assuming that the new nonstandard scalars decay through the SM particles only via couplings given in \Eqn{e:LNP}.} This limit is much above the
most conservative estimate for unitarity violation scale given by
\begin{eqnarray}
	\label{e:LamUV}
	\Lambda_{\rm UV}^{\rm max} =
	\sqrt{\frac{4\pi v^2}{\left|1-\kappa_W^h \kappa_Z^h\right|}} =
	\sqrt{\frac{4\pi v^2}{\left|2-\epsilon_W\right|}} \approx 650 ~{\rm GeV} \,.
\end{eqnarray} 
which is not substantially different from the bound in \Eqn{e:UVS}, obtained
for $\epsilon_W=0$. Thus, taking the experimental uncertainties in $\kappa_{W,f}^h$
is not sufficient to allow $\kappa_Z^h = -1$.

\begin{figure}[htbp!]
	\centering
		\includegraphics[scale=0.35]{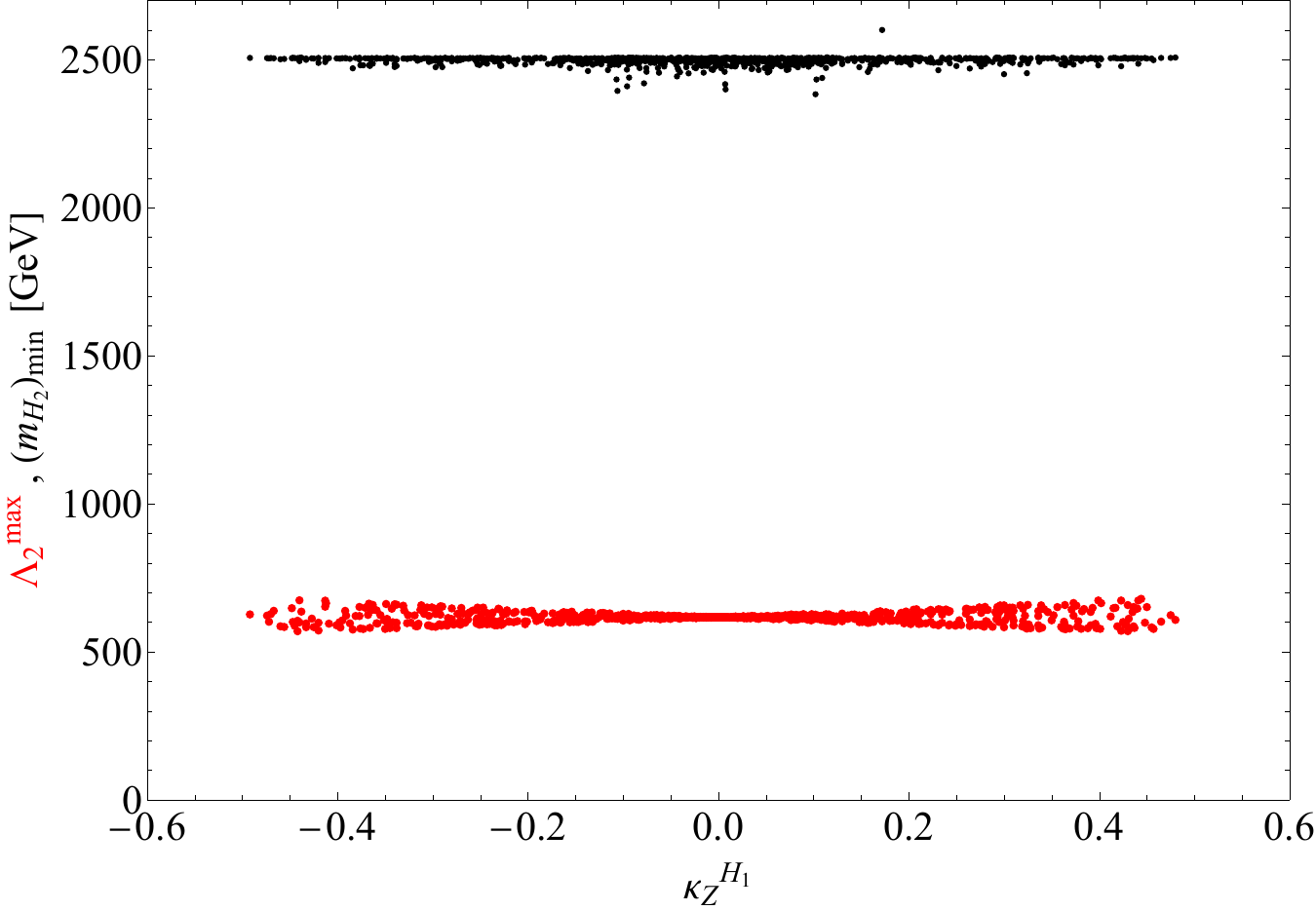}
	\caption{\it The unitarity violation scale ($\Lambda_2^{\rm max}$) for the second tier Higgs and the
		direct search limits on the second tier Higgs mass, $(m_{H_2})_{\rm min}$, plotted against
		$\kappa_Z^{H_1}$ as red and black regions respectively. We have used the bounds in
		Ref.~\cite{ATLAS:2023alo} to extract the experimental limits.
	}
	\label{f:2tier}
\end{figure}

As a final attempt, we consider the possibility of having multiple tiers of
BSM scalars which will ensure the unitarity of the theory at very high energies.
We illustrate this scenario by having two tiers of neutral BSM scalars ($H_1$
and $H_2$). The intent behind this strategy is to allot just enough coupling
strengths to the lighter BSM scalar~(say $H_1$) so that it can be allowed
before $\Lambda_1^{\rm max}\approx 620$~GeV (see \Eqn{e:UVS}) from direct
searches. After this, we hope that the assigned values of $\kappa_X^{H_1}$
will raise the second stage unitarity violation scale, $\Lambda_2^{\rm max}$,
 (dictating the onset
of effects of $H_2$) sufficiently so that $H_2$ can be comfortably accommodated
within $\Lambda_2^{\rm max}$ without being in conflict with the direct search
bounds.
Our analyses along these lines have been summarized in Fig.~\ref{f:2tier}. We have
first scanned $\kappa_X^{H_1}$ within the ranges\footnote{
The remaining parameters, namely, $\xi_1$ and $\kappa_{W,Z,f}^{H_2}$ can be solved
for using \Eqs{e:VVVV}{e:ffVV}.
}
\begin{eqnarray}
	\kappa_{W,Z,f}^{H_1} \in [-1, 1] \,, 
\end{eqnarray}
and collected the sets of values for which $H_1$ is allowed within $620$~GeV
from the direct searches. We have considered the search channels
$pp\xrightarrow{VBF}H_1 \to VV$ and $pp\xrightarrow{ggF} H_1 \to VV$
for this purpose. After this, we have
computed the upper limit on the mass of $H_2$ ($m_{H_2}$) using the formulae
for the second stage unitarity violation scales for the different scattering
amplitudes as follows:
\begin{subequations}
\label{e:Lam2}
	\begin{eqnarray}
	\Lambda_{WW\to ZZ}^{\rm max} &=& \sqrt{\frac{4\pi v^2}{\left|2-\kappa_W^{H_1} \kappa_Z^{H_1}\right|}}  \,,  \\
	\Lambda_{WW\to WW}^{\rm max} &=& \sqrt{\frac{8\pi v^2}{\left|\xi_1^2-(\kappa_W^{H_1})^2\right|}} \,,
	\end{eqnarray}
\end{subequations}
where $\kappa_f^h=\kappa_W^h= -\kappa_Z^h = 1$ has been already incorporated.
The lowest of the above is interpreted as $\Lambda_2^{\rm max}$ before which
the effects of $H_2$ must set in. The values of $\Lambda_2^{\rm max}$ estimated
using the sets of $\kappa_X^{H_1}$ and $\xi_1$ which successfully negotiate
the direct search constraints for $H_1$, have been plotted in Fig.~\ref{f:2tier}
as the red scattered region. The thickness of the plot arises due to the fact
that there are many possible values of $\kappa_{W,f}^{H_1}$ for a particular
value of $\kappa_Z^{H_1}$.

%
Next, we use these sets of values of $\kappa_X^{H_1}$ and $\xi_1$ that pass the
direct search limits on $H_1$ to compute the coupling strengths of $H_2$ ($\kappa_X^{H_2}$)
required to complete the unitarity sum rules of \Eqs{e:VVVV}{e:ffVV}. We then use
these values of $\kappa_X^{H_2}$ to compute the lower bounds on $m_{H_2}$ from
$pp\xrightarrow{VBF}H_2 \to VV$ and $pp\xrightarrow{ggF} H_2 \to VV$. The most
stringent of these lower bounds have been plotted as the black region in Fig.~\ref{f:2tier}.
As we can see from Fig.~\ref{f:2tier}, $(m_{H_2})_{\rm min}$ always lies above
$\Lambda_2^{\rm max}$ implying that the possibility of accommodating $\kappa_W^h
=-\kappa_Z^h =1$ with two tiers of neutral nonstandard scalars is also ruled out from
the combined constraints of unitarity and direct searches.
The reason behind this can be understood from the fact that rather small values
of $\kappa_X^{H_1}$ can be allowed (check the range on the horizontal axis
in Fig.~\ref{f:2tier}) in order to keep $m_{H_1}< 620$~GeV safe from
the direct searches. Consequently, the unitarity violation scale of \Eqn{e:UVS}
hardly gets relaxed for $\Lambda_2^{\rm max}$ as may be seen in the red region
in Fig.~\ref{f:2tier}. Therefore, for this strategy to work, one may need an
unusually large number of nonstandard scalars.

To summarize, we have proposed a complementary method to probe the sign
of the $hZZ$ coupling, which, unlike the existing studies, is immune to
the concerns regarding UV completion. Our approach is based on the realization
that the wrong-sign $hZZ$ coupling will bring in unitarity violation which
will necessitate existence of new particles with masses in the sub-TeV regime.
Moreover, the coupling strengths of these new particles can be inferred from
the unitarity sum rules. Thus, these new particles should be well within
the reach of the LHC and, if they are not observed in the direct searches,
the wrong-sign $hZZ$ coupling will be severely constrained. In fact, we have
been able to show that even with two tiers of nonstandard neutral scalars
brought in to restore unitarity, the wrong-sign $hZZ$ coupling cannot be
accommodated. Most importantly, our current analysis exemplifies the fact
that the so called {\em null} results in the direct searches can be translated
into nontrivial information regarding the couplings of the SM-like Higgs boson
thereby providing us with important insights about mechanism of electroweak
symmetry breaking. This may be treated as an additional incentive to carry
on the direct searches for new resonances at the current and the future
colliders.

\paragraph{Acknowledgements:}
DD thanks the Science and Engineering Research Board, India for financial support
through grant no. CRG/2022/000565.
AK acknowledges support from Science and Engineering Research Board, Government of India, through the grant CRG/2019/000362.
ML~acknowledges support from Fundação para a Ciência e a Tecnologia (FCT, Portugal) through the grant No.~PD/BD/150488/2019, in the framework of the Doctoral Programme IDPASC-PT, and from the projects~CERN/FIS-PAR/0002/2021, CFTP-FCT Unit~UIDB/00777/2020 and~UIDP/00777/2020, which are partially funded through POCTI (FEDER), COMPETE, QREN and EU.
AMP acknowledges support from UGC through the SRF fellowship scheme.
IS acknowledges the support from project number RF/23-24/1964/PH/NFIG/009073 and from DST-INSPIRE, India, under grant no. IFA21-PH272.
\bibliographystyle{JHEP}
\bibliography{hZZ_sign.bib}
\end{document}